\documentstyle[preprint,aps]{revtex}
\begin{document}
\draft
\preprint{IP/BBSR/94-1}
\title{GLUON CONDENSATES, QUARK MATTER EQUATION OF
STATE AND QUARK STARS}
\author{A. Mishra and H. Mishra}
\address{Theory Group, Physical Research Laboratory, Navrangpura, Ahmedabad
380 009, India}
\author{ P.K. Panda and S.P. Misra}
\address{Institute of Physics, Bhubaneswar-751005, India.}
\maketitle
\begin{abstract}
We consider here quark matter equation of state including strange quarks and
taking into account a nontrivial vacuum structure for QCD with gluon
condensates. The parameters of condendsate function are determined
from minimisation of the thermodynamic potential.
The scale parameter of the gluon condensates is fixed from the SVZ
parameter in the context of QCD sum rules at zero temperature
and zero baryon density. The equation of state for strange matter at
zero temperature as derived is used to study quark star structure using Tolman
Oppenheimer Volkoff equations. Stable solutions for quark stars
are obtained with a large Chandrasekhar limit as
3.2 $M_\odot$ and radii around 17 kms.
\end{abstract}
\narrowtext
\section{INTRODUCTION}
It is known that the vacuum structure of quantum chromodynamics
(QCD) is nontrivial and nonperturbative \cite{nielsen,svz,amqcd}.
One would like to
study the same in the context of field theory of quarks and
gluons. We believe that we may have hadronic to quark gluon
phase transition in heavy ion collisions. Signals from the
same however get masked since this phase lasts for a very short
time, and is followed by hadronisation. On the other hand, at
very high densities, as in neutron stars, we expect that hadrons
will dissolve to quark gluon matter. We wish to study the
same in the present paper with a nonperturbative equation
of state derived earlier \cite{zphyc}.

As a possible explanation for dark matter and otherwise,
Witten had suggested \cite{witten} that quark nuggets with
strange quarks may exist in the universe. Also, quark
stars as hypothetical objects have been considered
with perturbative QCD \cite{astro}. Hybrid stars
with the same perturbative approximation of QCD \cite{kapusta}
and semi-phenomenological treatment of hadronic matter
with a bag pressure has also been examined with various
types of models \cite{ellis,glendn,rosen}.
Recently, also hadronic matter had been considered
nonperturbatively with a perturbative treatment of
quark matter to study hybrid stars \cite{star}.
However, in a nonperturbative treatment
\cite{amqcd,zphyc} earlier, we found a change in the nature of
equation of state for quark matter.
We study here quark stars \cite{witten,astro} with the
above nonperturbative equation of state.
We find that some conclusions
differ by a factor as against perturbative calculations.

The equation of state for quark matter as will be used has been
discussed in Ref. \cite{zphyc}. We shall extend the same
including the strange quarks \cite{witten} and electrons with
charge neutrality condition.
We organise the paper as follows. In section II we shall
recapitulate the equation of state as was derived in
Ref. \cite{zphyc} along with modifications as above.
In section III we shall use this to
consider Tolman Oppenheimer Volkoff (TOV)  equations and
calculate different properties of hypothetical quark
stars. In section IV we summarise the results.

The motivation of the present analysis is to study bulk
properties in QCD with a
nonperturbative calculation. We find
the change in results from perturbative calculation to be significant.

\section {Equation of state for strange matter}
We had earlier considered the behaviour of gluon condensates at finite
baryon densities for up and down quarks \cite{zphyc}.
We shall here include the strange quark and electron
with charge neutrality condition to derive an
equation of state to study quark stars. For this purpose we
briefly recapitulate the notations of Ref. \cite{zphyc}
and extend it as above.
The QCD Lagrangian is given as
\begin{equation}
{\cal L} ={\cal L}_{gauge}+{\cal L}_{matter}+{\cal L}_{int},
\end{equation}
\noindent where
\begin{mathletters}
\begin{equation}
{\cal L}_{gauge}=-{1\over 2}G^{a\mu\nu}(\partial_{\mu}{W^{a}}_{\nu}
-\partial_{\nu}{W^{a}}_{\mu}+gf^{abc}{W^{b}}_{\mu}{W^{c}_{\nu})}
+{1\over 4}{G^{a}}_{\mu\nu}{G^{a\mu\nu}},
\end{equation}
\begin{equation}
{\cal L}_{matter}=\bar \psi (i \gamma ^\mu \partial _\mu-m_i)\psi
\end{equation}
\noindent and
\begin{equation}
{\cal L}_{int}=g\bar \psi \gamma ^\mu  \frac {\lambda^a}{2}
W _\mu^a\psi,
\end{equation}
\end{mathletters}
\noindent where ${W^{a}}_{\mu}$ are the SU(3) colour gauge fields.
We shall quantise in Coulomb gauge \cite{schwinger}
and write the
electric field  ${G^{a}}_{0i}$ in terms of
the transverse and longitudinal parts as
\begin{equation}
{G^a}_{0i}=
{^TG^a}_{0i}+{\partial_{i}{f}^{a}},
\end{equation}
where ${f}^{a}$ is to be determined.
We take at time t=0  \cite{amqcd}
\begin{mathletters}
\begin{equation}
{W^{a}}_{i}(\vec x)={(2\pi)^{-3/2}}\int
{d\vec k\over \sqrt{2\omega(\vec k)}}({a^a}_{i}(\vec k) +
{{a^a}_{i}(-\vec k)}^{\dagger})\exp({i\vec k.\vec x})
\end{equation}
\noindent and
\begin{equation}
{^{T}G^{a}}_{0i}(\vec x)={(2\pi)^{-3/2}} i \int
{d\vec k}{\sqrt{\omega(\vec k)\over 2}}(-{a^a}_{i}(\vec k) +
{{a^a}_{i}(-\vec k)}^{\dagger})\exp({i\vec k.\vec x}),
\end{equation}
\end{mathletters}
where, $\omega(k)$ is arbitrary \cite{schwinger}
and for equal time algebra we have
\begin{equation}
\left[ {a^a}_{i}(\vec k),{{a^b}_{j}(\vec k^{'})}^\dagger\right]=
\delta^{ab}\Delta_{ij}(\vec k)\delta({\vec k}-{\vec k^{'}}),
\end{equation}
with
\begin{equation}
\Delta_{ij}(\vec k)={\delta_{ij}}-{k_{i}k_{j}\over k^2}.
\end{equation}
The equal time quantization condition for the fermionic sector
is given as
\begin{equation}
[\psi _{\alpha}^i(\vec x,t),\psi _\beta^j (\vec y,t
)^{\dagger}]_{+}=\delta ^{ij}\delta _{\alpha \beta}\delta(\vec x -\vec y),
\end{equation}
where $i$ and $j$ refer to the colour and flavour indices
\cite{zphyc,pot}.
We now also have the field expansion for fermion field
$\psi$ at time t=0  given as \cite{zphyc,pot}
\begin{equation}
\psi^{i}(\vec x)=\frac {1}{(2\pi)^{3/2}}\int
\left[U_r(\vec k)c^{i}_{Ir}(\vec k)
+V_s(-\vec k)\tilde c^{i}_{Is}(-\vec k)\right]
e^{i\vec k\cdot \vec x} d\vec k,
\end{equation}
where $U$ and $V$ are given by \cite{spm78}
\begin{equation}
U_r(\vec k)=\left( \begin{array}{c}\cos \frac {\chi (k)}{2}
\\ \vec \sigma \cdot\hat k \sin \frac {\chi (k)}{2}
\end{array}\right)u_{Ir} ;\quad V_s(-\vec k)=\left(
\begin{array}{c}\vec \sigma .\hat k \sin \frac {\chi (k)}{2}
\\ \cos \frac {\chi(k)}{2} \end{array}\right)v_{Is},
\end{equation}
\noindent where the function $\chi(k)$ could be arbitrary.
Here we approximate the same as
$\cos \chi (k)=
{m_i}/{\epsilon_i (k)}$ and $\sin \chi(k)={k}/{\epsilon_i (k)}$,
with $\epsilon_i (k)=(k^2+m_i^2)^{1/2}$.
The above are consistent with the equal time anticommutation conditions
with \cite{spm78}
\begin{equation}
[c^{i}_{Ir}(\vec k),c^{j}_{Is}(\vec k')^\dagger]_{+}=
\delta _{rs}\delta^{ij}\delta(\vec k-\vec k')=
[\tilde c_{Ir}^{i}(\vec k),\tilde c_{Is}^{j}(\vec k')
^\dagger]_{+},
\end{equation}

In Coulomb gauge, the expression for the Hamiltonian
density, ${\cal T}^{00}$ from equation (1) is given as \cite{schwinger}
\begin{eqnarray}
{\cal T}^{00}&=&:{1\over 2}{^{T}{G^a}_{0i}}{^{T}{G^a}_{0i}}+
{1\over 2}{W^a}_{i}(-\vec \bigtriangledown^2){W^a}_{i}+
gf^{abc}W^a_iW^b_j\partial _i W^c_j\nonumber \\ &+&
{{g^2}\over 4}f^{abc}f^{aef}{W^b_i}{W^c_j}{W^e_i}{W^f_j}+
{1\over 2}(\partial_{i}f^{a})(\partial_{i}f^{a})
\nonumber \\ &+&\bar \psi (-i \gamma ^i \partial _i+m_i)\psi
-g\bar \psi \gamma ^ i \frac {\lambda^a}{2}
W _i^a\psi:,
\label{t00}
\end{eqnarray}
\noindent where : : denotes the normal ordering with respect to
the perturbative vacuum, say $\mid vac>$, defined through
${a^a}_{i}(\vec k)\mid vac>=0$, $c_{Ir}^i(\vec k)\mid vac>=0$
and $\tilde c_{Ir}^i(\vec k)^\dagger\mid vac >=0$.
In order to solve for the operator $f^a$, we first note that
\begin{equation}
f^a=-{W^a}_0-g \; f^{abc}\;{ (\vec \bigtriangledown ^2)}^{-1}
({W^b}_i \; \partial _i {W^c}_0).
\end{equation}
Proceeding as earlier \cite{amqcd,zphyc} with a mean
field type of approximation we obtain,
\begin{eqnarray}
\vec \bigtriangledown ^2{W^a}_0 (\vec x )
&&+ g^2 \; f^{abc}f^{cde} \;<vac',\beta\mid  {W^b}_i(\vec x ) \partial _i
(\vec \bigtriangledown ^2)^{-1}({W^d}_j(\vec x ) \mid vac',\beta>
\partial _j{W^e}_0(\vec x ))\nonumber\\ && ={J^a}_0(\vec x ),
\end{eqnarray}
where,
\begin{equation}
J^a_0=gf^{abc}{W^b_i}^{T}{G^c_{0i}}-g\bar \psi \gamma^0 \frac {\lambda ^a}{2}
\psi.
\end{equation}
In the above, $|vac',\beta>$ is the vacuum at finite temperature and
density in the language of thermofield dynamics \cite{tfd}.
We shall take the ansatz for the above state as \cite{zphyc}
\begin{equation}
|vac',\beta>=U_G(\beta)U_F(\beta)|vac'>,
\label{vacbeta}
\end{equation}
with $|vac'>$ being the gluon condensate vacuum at zero temperature
given as
\begin{equation}
\mid vac ^{'}>
=U\mid vac>\equiv
 \exp({B^\dagger}-B)\mid vac>
\end{equation}
obtained through the unitary operator $U$ on the pertubative
vacuum. In the above the condensate creation operator $B^\dagger $
is given explicitly as
\cite{zphyc}
\begin{equation}
{B^\dagger}={1\over 2}
\int {f(\vec k){{{a^a}_{i}(\vec k)}^\dagger}
{{{a^a}_{i}(-\vec k)}^{\dagger}}d\vec k},
\end{equation}
\noindent
where $f(\vec k)$ describes the vacuum structure with
gluon condensates. Further in equation (\ref {vacbeta})
 $U_G$ and $U_F$ are unitary operators \cite{tfd} involving
thermal excitations of gluons and fermions (quarks and
electrons) respectively.
For the gluon sector, we have \cite{amqcd,zphyc}
 \begin{equation}
U_G(\beta)
=\exp{({B_G(\beta)}^{\dagger}-B_G(\beta))},
\end{equation}
with
\begin{equation}
{B_G(\beta)}^{\dagger}=
\int \theta(\vec k,\beta){{b^a}_{i}(\vec k)}^{\dagger}
{{{\underline b}^a}_{i}(-\vec k)}^{\dagger}d\vec k.
\label{bgbeta}
\end{equation}
In addition, for fermionic sector we have,
\begin{equation}
U_F(\beta)
=\exp{({B_F(\beta)}^{\dagger}-B_F(\beta))},
\end{equation}
with
\begin{equation}
{B_F(\beta)}^{\dagger}=
\int \bigg[\theta^i_{-}(\vec k,\beta){{c^i}_{Ir}(\vec k)}^{\dagger}
{\underline c}^i_{Ir}(-\vec k)^{\dagger}
+ \theta^i_{+}(\vec k,\beta){\tilde c}^i_{Ir}(\vec k)
{\underline {\tilde c}^i}_{Ir}(-\vec k)
\bigg]\; d\vec k,
\label{bfbeta}
\end{equation}
where, $i$ refers to u, d, s quarks and electron. Further the
summation over the color and flavor indices  for the quarks
is understood and the underlined operators in equations
(\ref{bgbeta}), (\ref{bfbeta}) correspond to the extra Hilbert space
in thermofield dynamics.
$\theta(\vec k,\beta)$, $\theta^i_{\pm}(\vec k, \beta)$ are
functions related to the distribution functions and are given by \cite{tfd}
\begin{mathletters}
\begin{equation}
sinh^2\theta(\vec k,\beta) =\frac{1}{\exp(\beta\omega(\vec k,\beta))-1}
\label{distrb}
\end{equation}
and
\begin{equation}
sin^2\theta^i_{\pm}(\vec k,\beta) =\frac{1}{\exp(\beta(\epsilon^i
(\vec k)\pm \mu^i))+1}
\label{distrf}
\end{equation}
\end{mathletters}
where $\omega(k,\beta)=\sqrt{k^2+m_G^2}$,
$\epsilon^i(k)=\sqrt{k^2+m_i^2}$ and $\mu^i$ is the quark (or electron)
chemical potential.

Our job now is to evaluate the expectation value of
${\cal T}^{00}$ and then the thermodynamic potential
with respect to $\mid vac^{'};\beta>$.
The expressions obtained are almost the same as earlier
\cite{zphyc} except for additional contributions arising
from the electron and the strange quark sectors.
We thus have \cite{zphyc}
\begin{equation}
<vac';\beta \mid :{W^a}_{i}(\vec x){W^b}_{j}(\vec y):
\mid  vac';\beta >=
{\delta }^{ab}
\times (2  \pi )^{-3}\int d\vec k e^{i\vec k.(\vec x-
\vec y)}\; {F_{+}(\vec  k,\beta )\over \omega (k,\beta )}\;
\Delta _{ij}(\vec k),
\label{ww}
\end{equation}
\begin{equation}
{<vac';\beta \mid}: {^{T} G^{a}_{0i}} (\vec x)
{^{T} G^{b}_{0j}} (\vec y):{\mid vac';\beta >}
= \delta ^{ab}\times (2 \pi )^{-3}
\int d{\vec k}e^{i{\vec k}.{(\vec x-\vec y)}}
{\Delta _{ij}(\vec k)\omega (k,\beta )}
F_{-}( k,\beta ).
\label{gg}
\end{equation}
In the above the temperature dependant
$F_{\pm}(k,\beta )$ are given as
\begin{equation}
F_{\pm}(\vec k,\beta )  = \cosh 2\theta \bigg ({\sinh}^{2}f(k)
\pm{\sinh 2f(k)\over 2}\bigg ) +\sinh ^2 \theta (k,\beta )
\label{fpm}
\end{equation}
where $\sinh ^ 2 \theta (\vec k,\beta)$ is given by equation (\ref{distrb}).
Similarly, for the quark fields we have the parallel equations given as
\begin{mathletters}
\begin{equation}
<:\psi^{i}_\alpha(\vec x)^{\dagger}
\psi^{j}_\beta(\vec y):>_{vac',\beta}=
(2\pi)^{-3}\delta^{ij}\int
\Big ( \Lambda^i _-(\vec k,\beta)\Big )_{\beta\alpha}
e^{-i\vec k .(\vec x-\vec y)}d\vec k,
\label{jpj}
\end{equation}
\begin{equation}
<:\psi^{i}_\alpha(\vec x)
\psi^{j}_\beta(\vec y)^{\dagger}:>_{vac',\beta}=
(2\pi)^{-3}\delta^{ij}\int
\Big ( \Lambda^i _+(\vec k,\beta)\Big )_{\alpha \beta}
e^{i\vec k .(\vec x-\vec y)}d\vec k,
\label{jjp}
\end{equation}
\end{mathletters}
\noindent where
\begin{equation}
\Lambda^i_{\pm}(\vec  k,\beta)=\mp\frac{1}{2}\Big [\big
(\sin ^2 \theta _{-}^i-
\sin ^2 \theta _{+}^i\big )+\big (\gamma^0 \cos \chi+
\vec \alpha .\hat k \sin \chi\big)\big (\sin ^2 \theta^i_{-}
+\sin ^2 \theta^i_{+}\big )\Big ].
\end{equation}
Using equations (\ref{t00}), (\ref{ww}), (\ref{gg}) and (26),
we then obtain
the expectation value of ${\cal T}^{00}$ with respect to
$\mid vac^{'};\beta>$ as
\begin{eqnarray}
\epsilon_{0}(\beta) &
\equiv & <vac^{'};\beta \mid
:{\cal T}^{00}:\mid vac^{'};\beta> \nonumber \\
& = & C_{F}(\beta)+C_F^e (\beta)
+C_{1}(\beta)+C_{2}(\beta)+{C_{3}(\beta)}^{2}+C_{4}(\beta),
\label{enrgd}
\end{eqnarray}
\noindent where
\begin{mathletters}
\begin{eqnarray}
C_F(\beta)&=&<:\bar \psi (-i\gamma^i\partial _i +
m_i)\psi:>_{vac',\beta}\nonumber\\ &=&
\sum_{u,d,s}\frac{6}{\pi^2}\int {k^2dk}
(\sin^2 \theta^i_{-}(\vec k,\beta)+\sin^2 \theta^i_{+}(\vec k,\beta))
\epsilon^i(k),
\end{eqnarray}
\begin{eqnarray}
C_F^e(\beta)&=&<:\bar \psi (-i\gamma^i\partial _i
 )\psi:>_{vac',\beta}\nonumber\\ &=&
\frac{2}{\pi^2}\int {k^3dk}
(\sin^2 \theta^e_{-}(\vec k,\beta)+\sin^2 \theta^e_{+}(\vec k,\beta))
\end{eqnarray}
is the contribution for the electron,
\begin{eqnarray}
C_{1}(\beta) & = & <:{1\over 2}
{^T}{G^a}_{0i}{^T}{G^a}_{0i}:>_{vac^{'};\beta}\nonumber \\
& = & {4\over {\pi^2}}\int \omega(k,\beta)k^{2} F_{-}(k,\beta )\;dk,
\end{eqnarray}
\begin{eqnarray}
C_{2}(\beta ) & = & <:{1\over 2}
{W^a}_{i}{(-\vec \bigtriangledown^2)}{W^a}_{i}:>_{vac';\beta }\nonumber
\\ & = & {4\over {\pi^2}}\int {{k^{4}}\over \omega(k,\beta)}
\;F_{+}(k,\beta )\;dk
\end{eqnarray}
\begin{eqnarray}
{C_{3}(\beta )}^{2} & = & <:{1\over 4}g^{2}f^{abc}f^{aef}
{W^b}_{i}{W^c}_{j}{W^e}_{i}{W^f}_{j}:>_{vac';\beta }\nonumber \\
& = & \left({{2g}\over {\pi^2}}\int {{k^{2}}\over
{\omega(k,\beta )}}\;F_{+}(k,\beta )\;dk\right)^2 ,
\end{eqnarray}
\noindent and
\begin{eqnarray}
C_{4}(\beta ) & = & <:{1\over 2}
(\partial_{i}f^{a})(\partial_{i}f^{a}):>_{vac{'};\beta },\nonumber\\
 & = & 4\times (2 \pi )^{-6}\int  d \vec  k {G_1(\vec k,\beta )
+G_2(\vec k,\beta)\over
{k^2+\phi (k,\beta )}}.
\end{eqnarray}
\end{mathletters}
In the above,
\begin{mathletters}
\begin{eqnarray}
G_1(\vec k,\beta ) & = & 3 g^2 \int d  \vec q
F_{+}({\mid} \vec q\mid,\beta )\;
F_{-}({\mid} \vec k +\vec q {\mid},\beta  ) \;
{\omega ({\mid  \vec  k +\vec q \mid},\beta )\over
\omega  ({\mid \vec  q\mid},\beta  )}
\nonumber \\
& \times & \Bigl  (1+{{(q^2 +\vec k.\vec q)^2
}\over{q^2(\vec k+\vec q)^2}}\Bigr ),
\end{eqnarray}
and, now including the strange quark contribution,
\[G_2(\vec k, \beta)=\sum_{u,d,s} G^i_2(\vec k,\beta)\]
with
\begin{eqnarray}
G^i_2(\vec k,\beta) & = & -\frac{g^2}{2}\int d \vec q \Bigg [\Big (
1+\frac{m_i^2}{\epsilon_i (\vec q)\epsilon_i (\vec q-\vec k)}
+\frac {\vec q . (\vec q-\vec k)}
{\epsilon_i (\vec q)\epsilon_i (\vec q-\vec k)}\Big )\nonumber \\
& \times & \Big (\sin ^2 \theta^i_{-}(\vec q,\beta)\sin ^2 \theta^i_{-}(\vec
q-\vec k,\beta)
+\sin ^2 \theta^i_{+}(\vec q,\beta )\sin ^2 \theta^i_{+}(\vec q -\vec
k,\beta)\Big )
\nonumber\\ & - & \Big (
1-\frac{m_i^2}{\epsilon_i (\vec q)\epsilon_i (\vec q-\vec k)}
-\frac {\vec q . (\vec q-\vec k)}
{\epsilon_i (\vec q)\epsilon_i (\vec q-\vec k)}\Big )\nonumber \\
& \times &\Big (\sin ^2 \theta^i_{-}(\vec q,\beta)\sin ^2 \theta^i_{+}(\vec
q-\vec k,\beta)
+\sin ^2 \theta^i_{+}(\vec q,\beta )\sin ^2 \theta^i_{-}(\vec q -\vec
k,\beta)\Big )
\Bigg ]
\end{eqnarray}
\noindent  and
\begin{equation}
\phi  (k,\beta )=  {3g^2\over  {8 \pi  ^2}}
\int {{dk'}\over {\omega  (k',\beta  )}}\;F_{+}(k',\beta )
\biggl (  k^2+{k'}^2-{(k^2-{k'}^2)^2
\over{2kk'}}\log \Big | {{k+k'}\over{k-k'}}
\Big |  \biggr ).
\end{equation}
\end{mathletters}
As stated the expressions above are the same \cite{zphyc}
as earlier, except for the additional term
$C_F^e(\beta)$  as well as the additional contribution in
 $G_2(\vec k,\beta)$ in the expression for $C_4(\beta)$
arising from the strange quarks.

We shall take $\omega (\vec k,\beta)$ to be
of the free field form with a temperature dependent effective mass parameter
for the gluon fields given as
\begin{equation}
\omega (\vec k,\beta)=\sqrt {k^2+m_G(\beta)^2};\;\;
\end{equation}
\noindent with $m_G(\beta)$ given through the self consistency condition
\cite {amqcd,biro}
\begin{eqnarray}
m_G(\beta)^2 =
{{2g^2}\over {\pi^2}}\int {{k^{2}}\over
{\omega(k,\beta)}}\;F_{+}(k,\beta)\;dk\label{mg0}
\end{eqnarray}
from the single contraction  contribution from
the quartic terms \cite{zphyc}.

We are to now extremise over the thermodynamic potential
containing $\epsilon_0(\beta)$ including the additional
effects. For this purpose, we shall take the ansatz \cite{zphyc}
\begin{equation}
sinh f(\vec k)=Ae^{-Bk^2/2},
\end{equation}
which corresponds to taking a gaussian distribution for perturbative
gluons in nonperturbative vacuum \cite{amqcd}. The energy density,
$\epsilon_{0}(\beta)$ in terms of the dimensionless
quantities $x={\sqrt {B}}k$, $\mu_G={\sqrt B}m_G(\beta)$,
$m_i'={\sqrt B}m_i$, $\mu_i'={\sqrt B}\mu_i$
and $y={\beta\over {\sqrt B}}$ then gets parametrised as
\begin{eqnarray}
\epsilon_{0}(A,\beta)
& = & {1\over {B^2}}(I_F(y)+I_F^e(y)+I_{1}(A,y)+I_{2}(A,y)+{I_{3}(A,y)}^{2}+
I_{4}(A,y)) \nonumber \\
& \equiv & {1\over {B^2}}F(A,y),
\end{eqnarray}
\noindent where
\begin{mathletters}
\begin{equation}
I_F(y)=\sum_{u,d,s}{\frac{6}{\pi^2}\int {x^2dx}
(\sin ^2 \theta _{-}^i(\vec x,y)+\sin ^2 \theta _{+}^i(\vec x,y))
\epsilon^i(x)},
\end{equation}
\begin{equation}
I_F^e(y)=\frac{2}{\pi^2}\int {x^3dx}
(\sin ^2 \theta _{-}^e(\vec x,y)+\sin ^2 \theta _{+}^e(\vec x,y))
\end{equation}
\noindent  and
\begin{equation}
I_{4}(A,y) =4\times {(2 \pi )^{-6}}\int d{\vec x}
{{G_1(\vec x,y)+G_2(\vec x,y)}\over {x^2+\phi (x,y)}}.
\label{i4}
\end{equation}
\end{mathletters}
In the above, $G_1(\vec x,y)=G(\vec x,y)$ of \cite{amqcd}, and
\begin{eqnarray}
G_2(\vec x,y) & = & -\sum_{u,d,s}\frac{g^2}{2}\int d \vec x' \Bigg [\Big (
1+\frac{m'_i^2}{\epsilon_i (\vec x')\epsilon_i (\vec x'-\vec x)}
+\frac {\vec x' . (\vec x'-\vec x)}
{\epsilon_i (\vec x')\epsilon_i (\vec x'-\vec x)}\Big )\nonumber \\
& \times & \Big (\sin ^2 \theta^i_{-}(\vec x',y)
\sin ^2 \theta^i_{-}(\vec x'-\vec x,y)
+\sin ^2 \theta^i_{+}(\vec x',y )\sin ^2 \theta^i_{+}(\vec x' -\vec x,y)\Big )
\nonumber \\ & - &
\Big ( 1-\frac{m'_i^2}{\epsilon_i (\vec x')\epsilon_i (\vec x'-\vec x)}
-\frac {\vec x' . (\vec x'-\vec x)}
{\epsilon_i (\vec x')\epsilon_i (\vec x'-\vec x)}\Big )\nonumber\\&
 \times & \Big (\sin ^2 \theta^i_{-}(\vec x',y)
\sin ^2 \theta^i_{+}(\vec x'-\vec x,y)
+\sin ^2 \theta^i_{+}(\vec x',y )\sin ^2 \theta^i_{-}
(\vec x' -\vec x,y)\Big )
\Bigg ] .\nonumber\\& &
\end{eqnarray}
\noindent  The other expressions $I_1(A,y)$, $I_2(A,y)$, $I_3(A,y)$
and $\phi(x,y)$ in dimensional units are as earlier
\cite{amqcd,zphyc}. The above
integrals contain $\mu_G(y)$ which is
determined from the self consistency requirement \cite{amqcd}
\begin{eqnarray}
\mu_G(y)^2 & =& {{2g^2}\over {\pi^2}}
\int  {{x^2}dx\over \omega(x,y)}
\Biggl [ \biggl (A^{2}e^{-x^2}
+Ae^{-{{x^2}/2}}(1+A^{2}e^{-x^2})^{1/2}\biggr )
\biggl (1+{2\over \exp({y\omega(x,y)})-1}\biggr ) \nonumber \\ & + &
{1\over \exp({y\omega(x,y)})-1}\Biggr ],
\label{49}
\end{eqnarray}
The thermodynamic potential at temperature $T=1/\beta $ is given as \cite{tfd}
\begin{equation}
{\cal F}(A,\beta )=\epsilon _{0}{(A,\beta)}-{1\over \beta }(S_G+S_Q+S_e)
-\mu_B \rho_B - {\mu_E\rho_E}.
\label{free}
\end{equation}
Here $\mu_B$ is the baryon chemical poential corresponding to
the baryon number density $\rho_B$ given as,
\begin{equation}
\rho_B=\frac{1}{3}\times 3 \times 2\times\frac{1}{(2\pi)^3}
\int \Big (\sum_{u,d,s}\sin^2\theta^i_{-}- \sum_{u,d,s}\sin^2\theta^i_{+}\Big)
 d\vec k ,
\label{bnb}
\end{equation}
and $\mu^i$ as in equation (\ref{distrf}) is the quark chemical potential
for $i$th flavor. Further,
in the above, $\epsilon _{0}(A,\beta)$ is as given in equation (\ref{enrgd}),
and the entropy densities $S_G$, $S_Q$ and $S_e$ for the gluon, quark and
electron fields
are given as \cite{tfd}
\begin{equation}
S_G=-2\times 8 \times (2\pi)^{-3}
\int d{\vec k}\Big (\sinh ^{2} \theta \; log \bigl(\sinh ^{2} \theta\bigr)
-\cosh ^{2} \theta \; log \bigl(\cosh ^{2} \theta\bigr) \Bigr ),
\label{sg}\end{equation}
\begin{eqnarray}
S_Q&=&-\sum_{u,d,s}\bigg[3\times 2\times (2\pi)^{-3}
\int d{\vec k}\Big (\sin ^2 \theta _{-}^i(\vec k,\beta) log
\big(\sin ^2 \theta _{-}^i(\vec k,\beta)\big)\nonumber\\
&+&(\cos ^2 \theta _{-}^i(\vec k,\beta))
log  \big(\cos ^2 \theta _{-}^i(\vec k,\beta)\big)
+\sin ^2 \theta _{+}^i(\vec k,\beta)
log  \big(\sin ^2 \theta _{+}^i(\vec k,\beta)\big)\nonumber\\
&+&(\cos ^2 \theta _{+}^i(\vec k,\beta)) log
\big(\cos ^2 \theta _{+}^i(\vec k,\beta)\big)\Big )\big] .
\label{sf}
\end{eqnarray}
and
\begin{eqnarray}
S_e&=&-2\times (2\pi)^{-3} \int d{\vec k}
\Big (\sin ^2 \theta _{-}^e(\vec k,\beta) log
\big(\sin ^2 \theta _{-}^e(\vec k,\beta)\big)\nonumber\\
&+&(\cos ^2 \theta _{-}^e(\vec k,\beta))
log  \big(\cos ^2 \theta _{-}^e(\vec k,\beta)\big)
+\sin ^2 \theta _{+}^e(\vec k,\beta)
log  \big(\sin ^2 \theta _{+}^e(\vec k,\beta)\big)\nonumber\\
&+&(\cos ^2 \theta _{+}^e(\vec k,\beta)) log
\big(\cos ^2 \theta _{+}^e(\vec k,\beta)\big)\Big )\big] .
\label{se}
\end{eqnarray}
Clearly, the factor $2\times 8$ in equation
(\ref{sg}) above comes from the transverse
and colour degrees of freedom for the gluon fields
and the factor $3\times 2$ in  equation (\ref{sf}) comes from the
colour and spin degrees of freedom for quarks and the factor 2
in equation (\ref{se}) comes from spin degrees of freedom for electrons.
We may note that to consider quark stars we need to impose charge neutrality
condition for the quark matter. The electric charge $\rho_E$ is given by
\begin{eqnarray}
\rho_E & = & \frac{2}{(2\pi)^3}\int
d\vec k \bigg[ 2(\sin^2\theta^u_-  - sin^2\theta^u_+)
-(\sin^2\theta^d_- - sin^2\theta^d_+)\nonumber \\
& - & (\sin^2\theta^s_- - sin^2\theta^s_+)-(\sin^2\theta^e_- -
sin^2\theta^e_+) \bigg] ,
\label{rhoe}
\end{eqnarray}
where $\sin^2\theta_{\mp}^e$ is the electron (positron)
distribution function parallel to equation (\ref {distrf}). The chemical
potentials for the quarks and the electrons can be written
in terms of baryon chemical potential $\mu_B$ and the electron chemical
potential $\mu_E$ and are given by
\begin{mathletters}
\begin{equation}
\mu_u=\frac{1}{3}\mu_B+\frac{2}{3}\mu_E
\label{mua}
\end{equation}
\begin{equation}
\mu_d=\frac{1}{3}\mu_B-\frac{1}{3}\mu_E
\label{mub}
\end{equation}
\begin{equation}
\mu_s=\frac{1}{3}\mu_B-\frac{1}{3}\mu_E
\label{muc}
\end{equation}
\begin{equation}
\mu_e=-\mu_E
\end{equation}
\label{mud}
\end{mathletters}
For the minimisation of thermodymic potential we may scale out the
dimensional parameter $1/{B^2}$ as in Ref. \cite{zphyc} and have parallel
expression for the thermodynamic potential as
\begin{eqnarray}
{\cal F}(A,\beta )& \equiv &  {1\over {B^2}}{F_1}(A,y)\nonumber
\\ & = &
{1\over {B^2}}\bigg [F(A,y)-{1\over y}({\cal S}_G(A,y)
+{\cal S}_Q(A,y)+{\cal S}_e(A,y))-\mu_B ' \rho_B '\bigg ],
\label{52}
\end{eqnarray}
\noindent where $F(A,y)=B^2 \epsilon _0(A,\beta)$, $\mu_B '=\sqrt{B}\mu_B$,
 $\rho_B '= B^{3/2}\rho_B$
and the entropy densities ${\cal S}_G(A,y)$, ${\cal S}_Q(A,y)$
and ${\cal S}_e(A,y)$ in dimensionless units are given as
\begin{eqnarray}
{\cal S}_G(A,y)&=&-{8\over {\pi^2}}  \int  x^{2}dx \biggl \{
\bigl ({1\over \exp{(y\omega(x,y))}-1}\bigr )
log \bigl ({1\over \exp{(y\omega(x,y))}-1}\bigr )\nonumber\\ & - &
\bigl (1+{1\over \exp{(y\omega(x,y))}-1}\bigr )
log \bigl (1+{1\over \exp{(y\omega(x,y))}-1}\bigr )\biggr \},
\end{eqnarray}
\begin{eqnarray}
{\cal S}_Q(A,y)&=&\sum_{u,d,s}-{6\over {\pi^2}} \int x^{2}dx \biggl \{
\bigl ({1\over \exp{(y(\epsilon_i(x)-\mu'_i)}+1}\bigr )
log \bigl ({1\over \exp{(y(\epsilon_i(x)-\mu'_i)}+1}\bigr )\nonumber\\ & + &
\bigl (1-{1\over \exp{(y(\epsilon_i(x)-\mu'_i)}+1}\bigr )
log \bigl (1-{1\over \exp{(y(\epsilon_i(x)-\mu'_i)}+1}\bigr )
\nonumber\\ & + & \bigl ({1\over \exp{(y(\epsilon_i(x)+\mu'_i)}+1}\bigr )
log \bigl ({1\over \exp{(y(\epsilon_i(x)+\mu'_i)}+1}\bigr )\nonumber\\ & + &
\bigl (1-{1\over \exp{(y(\epsilon_i(x)+\mu'_i)}+1}\bigr )
log \bigl (1-{1\over \exp{(y(\epsilon_i(x)+\mu'_i)}+1}\bigr )
\biggr \},
\label{mu}
\end{eqnarray}
and
\begin{eqnarray}
{\cal S}_e(A,y)&=&-{2\over {\pi^2}} \int x^{2}dx \biggl \{
\bigl ({1\over \exp{(y(\epsilon_e(x)-\mu'_e)}+1}\bigr )
log \bigl ({1\over \exp{(y(\epsilon_e(x)-\mu'_e)}+1}\bigr )\nonumber\\ & + &
\bigl (1-{1\over \exp{(y(\epsilon_e(x)-\mu'_e)}+1}\bigr )
log \bigl (1-{1\over \exp{(y(\epsilon_e(x)-\mu'_e)}+1}\bigr )
\nonumber\\ & + & \bigl ({1\over \exp{(y(\epsilon_e(x)+\mu'_e)}+1}\bigr )
log \bigl ({1\over \exp{(y(\epsilon_e(x)+\mu'_e)}+1}\bigr )\nonumber\\ & + &
\bigl (1-{1\over \exp{(y(\epsilon_e(x)+\mu'_e)}+1}\bigr )
log \bigl (1-{1\over \exp{(y(\epsilon_e(x)+\mu'_e)}+1}\bigr )
\biggr \}.
\label{mue}
\end{eqnarray}
We note that for each $A$, the gluon mass is
determined self consistently through equations (38)
for the evaluation of the right hand side of equation (46). As has
already been stated, unlike in Ref. \cite{zphyc} we have now to impose
charge neutrality condition i.e., $\rho_E=0$ to discuss quark star.
Thus, for a given baryon density $\rho_B$, the chemical potentials
$\mu_B$ and $\mu_E$ are fixed using equations (\ref{bnb})  and (\ref{rhoe})
demanding charge neutrality. This with equations (44)
fixes up the chemical potential of each species of fermions.
With the chemical potentials known, we now extremise the
thermodynamic potential
$F_{1}(A,y)$  of equation (\ref{52}) with
respect to the parameter A and obtain the optimum value of A as $A_{min}$ at
a given temperature T and baryon density $\rho_B$.
We then calculate the pressure $P$ given as \cite{fetter}
\begin{equation}
P(\beta )=-{\cal F}(A,\beta )\Big |_{A=A_{min}}.
\end{equation}

We calculate the pressure with respect to the
nonperturbative vacuum as a function of the condensate
parameter $A$ for different baryon densities $\rho_B$ for coupling constant
equal to $0.8$ and for zero temperature. We have also taken here current quark
masses of u, d quarks to be zero and starnge quark mass to be 180 MeV. We
may mention here that the dimensional parameter 1/$\sqrt B$ for coupling
$\alpha_s=0.8$ is 321.4 MeV which is fixed by relating it to the SVZ parameter
$<{\alpha_s\over\pi}G^{ a}_{\mu\nu}^aG^{\mu\nu}>=0.012 GeV^4$
at zero temperature and zero density \cite{amqcd,zphyc}. This is used to
express different thermodynamic quantities in physical units. At
zero density the pressure has a unique minimum at $A=A_{min}=0.56$.
For density greater than a critical density
$\rho_{crit}=0.55\;$baryons/fm$^3$, the pressure is maximum for
A=0 or, equivalently gluon condensate vanishes.

We plot in Fig. 1 pressure versus baryon density at zero temperature
after subtracting the nonperturbative $"$bag constant" here
calculated as 38.3 MeV/fm$^3\equiv$ (131 MeV)$^4$.
We may compare the present nonperturbative results
with pressure obtained from pertubative calculation \cite{kapusta}
which is given as
\begin{equation}
P_{pert}=P_e+\sum_{u,d,s} P_f,
\end{equation}
with the electron pressure $P_e$ as
\begin{equation}
P_e=\frac{\mu_e^4}{12\pi^2}
\end{equation}
and
\begin{eqnarray}
P_f& = & \frac{1}{4\pi^2}\bigg [ \mu_f k_f(\mu_f^2-2.5m_f^2)
+1.5 m_f^4 ln \left (\frac {\mu_f+k_f}{m_f}\right ) \bigg]\nonumber\\
& - & \frac{\alpha_s}{\pi^3}\bigg [1.5\times\left (\mu_f k_f -m_f^2
ln \big (\frac {\mu_f+k_f}{m_f} \big )\right )^2-k_f^4\bigg ],
\label{pert}
\end{eqnarray}
where $k_f=\sqrt{\mu_f^2-m_f^2}$ with $\mu_f$ being the
chemical potential for quark of flavor $f$.
As may be seen, the present equation of state is stiffer than
the perturbative equation of state
which is shown as a dashed curve in Fig 1.
Pressure as a function of baryon chemical potential is
plotted in Fig. 2.
\section{quark stars}
We shall use the equation of state for the
strange quark matter as obtained here
for the consideration of the quark stars.
For the description of a spherical strange quark star,
the space time geometry is described by a metric in
Schwarzchild coordinates and has the form \cite{weinberg}
\begin{equation}
ds^2=-e^{\nu(r)}dt^2+[1-2M(r)/r]^{-1}dr^2+r^2[d\Theta^2+sin^2 \Theta
d\phi^2]
\end{equation}
The equations which determine the star structure and the geometry are,
in dimensionless forms \cite{weinberg},
\begin{mathletters}
\begin{equation}
{d\hat P(\hat r r_0)\over d\hat r}=-\hat G
{[\hat\epsilon (\hat r r_0)+\hat P (\hat r r_0)][\hat M (\hat r r_0)
+4\pi a \hat r^3 \hat P(\hat r r_0)]\over \hat r^2[1-2\hat G
\hat M (\hat r r_0)/\hat r]},\label{tov1}
\end{equation}
\begin{equation}
\hat M (\hat r r_0)=4\pi a \int_0^{\hat r r_0} d\hat r^\prime
\hat r^{\prime^2} \hat \epsilon(\hat r^\prime r_0),\label{tov2}
\end{equation}
and the metric function, $\nu (r)$, relating the element of time at
$r=\infty$ is given by
\begin{equation}
{d\nu(\hat r r_0)\over d\hat r}=2\hat G {[\hat M (\hat r r_0)
+4\pi a \hat r^3 \hat P(\hat r r_0)]\over \hat r^2[1-2\hat G
\hat M (\hat r r_0)/\hat r]}.\label{tov3}
\end{equation}
\end{mathletters}
In equations (54) the following substitutions have been made.
\begin{mathletters}
\begin{equation}
\hat \epsilon\equiv \epsilon/\epsilon_c,\quad
\hat P\equiv P/\epsilon_c,\quad\hat r\equiv r/r_0,\quad
\hat M\equiv M/M_\odot,
\label{tov4}
\end{equation}
where, with
$ f_1=197.3 \; MeV^{-1}/fm^3$ and $r_0=3\times 10^{19}$ fm,
we have
\begin{equation}
a\equiv\epsilon_c r_0^3/M_\odot, \quad
\hat G\equiv (G/f_1)/(r_0/M_\odot)
\label{tov5}
\end{equation}
\end{mathletters}
\noindent In the above, quantities with hats are dimentionless.
G in equation (\ref{tov5}) denotes the gravitational constant
$(G=6.707934\times 10^{-45}\;\mbox{MeV}^{-2})$.

In order to construct a stellar model, one has to integrate equations
(\ref{tov1}) to  (\ref{tov3}) from the star's center at r=0 with
a given central energy density $\epsilon_c$ as input until the
pressure $P(r)$ at the surface $"$vanishes", i.e., becomes the
same as the pressure for the nonperturbative vacuum, or bag pressure.
Thus we shall here use equation of state of quark matter through
equation (49) after subtracting from it the bag pressure,
which is the difference in energy density of the nonperturbative
vacuum and the perturbative vacuum.
We then integrate the TOV equations
with $\hat P (0)= P (\epsilon_c)$.
As mentioned in the last section, for $\rho > \rho_{crit}$, the
condensate structure vanishes. For TOV equations, we use the
corresponding equation of state in the range $\rho_{crit}<\rho
<\rho_C$, i.e.,
until the density decreases to the critical value $\rho_{crit}$.
We then use the equation of state when
the gluon condensates exist. TOV equations continue
till the pressure vanishes, which defines the surface of the star.
This completes the calculations for stellar model for quark stars
whose mass and radius could be calculated for different central densities.

In Fig. 3 we plot the mass of a star as a function of central energy density to
examine the stability of such stars. As may be seen from the figure
$dM/d\epsilon$ becomes negative around central densities 950 MeV/fm$^3$
after which they may collapse to black holes \cite{weinberg,shapiro} with the
maximum mass as 3.2$M_\odot$, which is rather large.
We have also made a calculation, not shown in the present
figures, for the critical mass with the perturbative equation
of state taking the same coupling constant. We then obtain that
$M_{max}\simeq$ 2.25$M_\odot$. Since the equation of state
for strange quark matter as obtained here is stiffer than the
perturbative equation of state as in Fig. 1, a larger value
of $M_{max}$ is expected \cite{shapiro,prakash}.
A similar large mass for hybrid stars was obtained in an
effective condensate model for QCD \cite{ellis}.

Fig. 4 shows the mass as a function of radius  for such stars obtained
for different central densities varying in the range
500 to 1200 MeV/fm$^3$. This yields stable quark stars of masses $M\simeq$
2.2 to 3.2 $M_\odot$ with radii $R\simeq$ 15.3 to 17 kms respectively.

The energy density profile obtained from (\ref{tov1}) to (\ref{tov3}) is
plotted in Fig. 5 for central density $\epsilon_c=850$ MeV/fm$^3$.

We may also calculate the surface gravitational red shift $Z_s$ of
photons which is given by \cite{weinberg,brecher}
\begin{equation}
Z_s={1 \over\sqrt{[1-2GM/R]}} -1.\label{zs}
\end{equation}
In Fig 6 we plot $Z_s$ as a function of $M/M_{\odot}$.
\section{conclusions}
Let us summarise the findings of the present results.
The nontrivial ground state structure of QCD with gluon condensates
at zero temperature \cite{amqcd} has been generalised here to
finite baryon densities with three quark flavour.
We have here included the charge neutrality condition
to obtain the equation of state for strange quark matter.
The equation of state so obtained is seen
to be stiffer than the perurbative equation of state.
The gluon condensate appears to vanish
for baryon density greater than a critical value.

To derive the equation of state we have however not included the effects
of vacuum structure with quark condensates associated with
chiral symmetry breaking, the reason for this being that
the contribution of quark condensates to the energy density
is expected to be small compared to that of the gluons. However
inclusion of the same might throw a light on relationship
between confinement and chiral symmetry breaking.

We have applied the equation of state so obtained to study
structural properties of quark stars. They seem to be larger and
in general more massive than neutron stars or hybrid stars with masses
being around $2.2M_\odot$ to $3.2M_\odot$
and radii around 15 to 17 kilometers.
We may note that this is a reflection of a stiffer equation
of state of strange quark matter as obtained here.

\acknowledgements

The authors are thankful to J.C. Parikh, S.B. Khadkikar, M. Prakash,
N. Barik and Snigdha Mishra for many useful discussions.
SPM would like to thank Department of
Science and Technology, Government of India for
research grant no SP/S2/K-45/89 for financial assistance.

\def \qcd {G. K. Savvidy, Phys. Lett. 71B, 133 (1977);
S. G. Matinyan and G. K. Savvidy, Nucl. Phys. B134, 539 (1978); N. K. Nielsen
and P. Olesen, Nucl.  Phys. B144, 376 (1978); T. H. Hansson, K. Johnson,
C. Peterson Phys. Rev. D26, 2069 (1982).}

\def \svz {M.A. Shifman, A.I. Vainshtein and V.I. Zakharov,
Nucl. Phys. B147, 385, 448 and 519 (1979);
R.A. Bertlmann, Acta Physica Austriaca 53, 305 (1981).}

\def \nambu{ Y. Nambu, Phys. Rev. Lett. 4, 380 (1960);
Y. Nambu and G. Jona-Lasinio, Phys. Rev. 122, 345 (1961); ibid,
124, 246 (1961);
J.R. Finger and J.E. Mandula, Nucl.Phys.B199, 168 (1982);
A. Amer, A. Le Yaouanc, L. Oliver, O. Pene and
J.C. Raynal, Phys. Rev. Lett. 50, 87 (1983);
ibid, Phys. Rev. D28, 1530 (1983);
 S.L. Adler and A.C. Davis,
Nucl. Phys. B244, 469 (1984); R. Alkofer and P. A. Amundsen,
Nucl. Phys.B306, 305 (1988); A.C. Davis and A.M. Matheson DAMTP 91-34 (1991).}

\def \pot{A. Mishra, H. Mishra and S. P. Misra, Z. Phys. C57, 241 (1993);
A. Mishra and S. P. Misra, Z. Phys. C58, 325 (1993).}

\def \lat{K.G. Wilson, Phys. Rev. D10, 2445 (1974); J.B. Kogut, Rev. Mod.
Phys. 51, 659 (1979); ibid 55, 775 (1983); M. Cruetz, Phys. Rev. Lett.
45, 313 (1980); ibid Phys. Rev. D21, 2308 (1980); T. Celik, J. Engels and
H. Satz, Phys. Lett. B129, 323 (1983); H. Satz, in Proceedings of Large Hadron
Collider Workshop, Vol.I, Ed. G. Jarlskog and D. Rein, CERN 90-10, ECFA 90-133
(1990).}

\def \biroo {T. Biro, Ann. Phys. 191, 1 (1989), Phys. Lett B228, 16 (1989);
Phys. Lett. B245, 142 (1990).}
\def \joglekar {J. Collins, A. Duncan and S. Joglekar, Phys. Rev.
D16, 438 (1977); N.K. Nielsen, Nucl. Phys. B120, 212 (1977);
J. Schechter, Phys. Rev. D21, 3393 (1981); A.A. Migdal
and M.A. Shifman, Phys. Lett. 114B, 445 (1982).}

\def \star {H. Mishra, S.P. Misra, P.K. Panda and B.K. Parida,
to appear in Int. J. Mod. Phys. E.; H. Mishra,
 S.P. Misra, P.K. Panda and B.K. Parida, Int. J. Mod. Phys. E1, 405 (1992).}

\def \amqcd { A. Mishra, H. Mishra, S.P. Misra and S.N. Nayak,
Pramana (J. of Phys.) 37, 59 (1991); A. Mishra, H. Mishra, S.P. Misra
and S.N. Nayak, Z. Phys. C57, 233 (1993).}

\def \tfd {H. Umezawa, H. Matsumoto and M. Tachiki,{\it
Thermofield Dynamics and Condensed States} (North Holland,
Amsterdam, 1982).}

\def \nmtr {H. Mishra,
 S.P. Misra, P.K. Panda and B.K. Parida, Int. J. Mod. Phys. E1, 405 (1992).}

\def \schwinger{  J. Schwinger, Phys. Rev. 125, 1043 (1962); ibid,
127, 324 (1962); E. S. Abers and B. W. Lee, Phys. Rep. 9C, 1 (1973);
D. Schutte, Phys. Rev. D31, 810 (1985).}

\def \spm { S. P. Misra, Phys. Rev. D18, 1661, 1673 (1978);
A. Le Youanc et al Phys. Rev. Lett. 54, 506
(1985).}

\def \hmgrnv { H. Mishra, S.P. Misra and A. Mishra,
Int. J. Mod. Phys. A3, 2331 (1988);
S.P. Misra, Phys. Rev. D35, 2607 (1987).}

\def \fetter {A. L. Fetter and J. D. Walecka, {\it Quantum Theory of
Many Particle Systems}, McGraw Hill Book Company, 1971.}

\def \kapusta {J.I. Kapusta, {\em Finite Temperature
Field Theory }(Cambridge University Press 1989);
B. Freedman and L. McLerran, Phys. Rev. D17, 1109
(1978); B. D. Serot and H. Uechi, Ann. Phys. 179, 272 (1987).}

\def \biro {Tamas S Biro, Int. J. Mod. Phys. E1, 39 (1992).}

\def \satz {H. Satz, Nucl. Phys. A498, 495c (1989).}

\def \greiner {P.R. Subramanian, H. Stocker and W. Greiner,
Phys. Lett. B173,  468 (1986).}
\def \quenched {D. Barkai, K.J.M. Moriarty and C. Rebbi,
Phys. Rev. D30, 1293 (1984); A.D. Kennedy, J. Kuti, S.Meyer and B.J. Pendelton,
Phys. Rev. Lett. 54, 87 (1985).}
\def \unquenched {S. Gottlieb et al, Phys. Rev. D38, 2245 (1988).}
\def \gavai {R.V. Gavai in Proceedings of Workshop on Role of Quark
Matter In Physics and Astrophysics, edited by R.S. Bhalerao and R.V. Gavai,
Bombay 1992.}
\def \zphyc {A. Mishra, H. Mishra and S.P. Misra, Z. Phys. C59, 159 (1993).}
\def \witten{ E. Witten, Phys. Rev. D30, 272 (1984);
E. Farhi and R. L. Jaffe, Phys. Rev. D30, 2379 (1984).}
\def \glendn {N.K. Glendenning, Talk presented at the Workshop on
Strange Quark Matter in Physics and Astrophysics, Aarhus, Denmark, 1991, to
appear in the proceedings; ibid preprint LBL-30878.}
\def \ellis{J. Ellis, J.I. Kapusta and K.A. Olive, Nucl. Phys. B348,
345 (1991); ibid, Phys. Lett. B273, 123 (1991); G. E. Brown and Mannque Rho,
Phys. Rev. Lett. 66, 2720 (1991).}
\def \astro {T. \O verg\.{a}rd and E. \O verg\.{a}rd,
Astro. Astrophys. 243, 412 (1991); ibid, Class.
Quantum Grav. 8, L49, 1991; B.Datta, P. K. Sahu, J. D. Anand
and A. Goyal, Phys. Lett. B283, 313 (1992).}
\def\rosen {A. Rosenhauer et al  Nucl. Phys. A540, 630 (1992).}
\def\weinberg {S. Weinberg, {\it Gravitation and cosmology}, (Wiley,
New York, 1972);
F. Weber and M.K. Weigel, Nucl. Phys. {\bf A493} (1989) 549.}
\def\shapiro {S. L. Shapiro and S. A. Teulosky, {\it Black holes,
white dwarfs and neutron stars} (Wiely, New York, 1983).}
\def\prakash {J. M. Lattimer, M. Prakash, D. Masak
and A. Yahil, Astro. Phys. J. 355 (1990) 241; V. Thorsson,
M. Prakash and J. M. Lattimer, NORDITA-93/29 N, SUNY-NTG-92-33.}
\def\brecher {K. Brecher, Astro. Phys. J. {\bf 215} (1977) L17.}

\newpage
\centerline{\bf Figure Captions}
\bigskip
\noindent {\bf Fig.1:} We plot pressure, P in MeV/fm$^3$
as a function of the baryon number density, $\rho_B$
in fm$^{-3}$.
The dashed curve is the perturbative equation of state.
\hfil
\medskip

\noindent {\bf Fig.2:} We plot here pressure as a function of
baryon chemical potential $\mu_B$.
\hfil
\medskip

\noindent {\bf Fig.3:} Mass of quark star as a function
of central density is plotted here. The dashed portion
indicates instability.
\hfil
\medskip

\noindent {\bf Fig.4:} We plot here mass of the star in units of solar mass
as a function of radius of the star in kilometers.
\hfil
\medskip

\noindent {\bf Fig.5:} We plot  here the energy profile
inside the quark star for central density 850 MeV/fm$^3$.
\hfil
\medskip

\noindent {\bf Fig.6:} We plot  here the surface gravitational
red shift as a function of $M/M_\odot$.
\hfil
\medskip
\end{document}